\documentclass[%
 reprint,
 amsmath,amssymb,
 aps,
]{revtex4-2}

\usepackage{graphicx}
\usepackage{dcolumn}
\usepackage{bm}
\usepackage{hyperref}
\usepackage[mathlines]{lineno}
\usepackage{makecell}
\usepackage{color}
\usepackage{multirow}
\usepackage{booktabs}
\usepackage{amsmath}
\usepackage{amssymb}
\usepackage{amsthm}
\usepackage{bbm}
\usepackage{hyperref}
\usepackage[T1]{fontenc}
\usepackage[usenames,dvipsnames]{xcolor}
\hypersetup{colorlinks=true, linkcolor=blue, urlcolor=blue, citecolor=blue}

\begin{document}

\preprint{APS/123-QED}

\title{Time-dependent invasion laws for a liquid-liquid displacement system}

\author{Ke Xiao}
\email{xiaoke@ucas.ac.cn}
 \affiliation{Department of Physics, College of Physical Science and Technology, Xiamen University, Xiamen 361005, People's Republic of China}
\author{Chen-Xu Wu}%
 \email{cxwu@xmu.edu.cn}
\affiliation{Fujian Provincial Key Lab for Soft Functional Materials Research, Research Institute for Biomimetics and Soft Matter, Department of Physics, College of Physical Science and Technology, Xiamen University, Xiamen 361005, People's Republic of China}

\date{\today}

\begin{abstract}
Capillary-driven flow of fluids occurs frequently in nature and has a wide range of technological applications in the fields of industry, agriculture, medicine, biotechnology, and microfluidics. By using the Onsager variational principle, we propose a model to systematically study the capillary imbibition in titled tubes, and find different laws of time-dependent capillary invasion length for liquid-liquid displacement system other than Lucas-Washburn type under different circumstances. The good agreement between our model and experimental results shows that the imbibition dynamics in a capillary tube with a prefilled liquid slug can be well captured by the dynamic equation derived in this paper. Our results bear important implications for macroscopic descriptions of multiphase flows in microfluidic systems and porous media.
\end{abstract}

\maketitle


\section{\label{sec:introduction}INTRODUCTION}
The capillary rise phenomenon is one of the most well-known and vivid illustrations of capillarity owing to its relevance to numerous natural and industrial processes such as liquid transport in porous media~\cite{N.R.Morrow1970,N.Lu2004,T.Dang-Vu2005,S.Gruener2009}, wetting of fabrics~\cite{F.Ferrero2003,C.Duprat2022}, additive manufacturing~\cite{G.Arrabito2019,A.Azhari2017}, and microfluidic devices for liquid handling~\cite{C.Li2019,C.E.Anderson2022}.
Such a capillary flow in a confined environment has a wide range of technological applications, such as brain capillary flow~\cite{A.A.Berthiaume2022}, microfluidic diagnostics~\cite{D.Shou2018}, lab-on-a-chip devices~\cite{A.Olanrewaju2018,J.Park2020}, fabrication of flexible printed electronics~\cite{M.Cao2018,K.S.Jochem2018}, and oil recovery from porous reservoirs~\cite{N.R.Morrow2001,B.A.Suleimanov2011}.
It concerns a spontaneous elevation of liquid imbibed into a capillary tube without the assistance of external forces when inserted into a bath of liquid~\cite{J.Cai2021,P.Kolliopoulos2021}. The classical celebrated Lucas-Washburm (LW) law reveals that the invasion length of the advanced meniscus ($x$) usually follows a diffusive-like scaling behavior of $x\propto t^{1/2}$~\cite{R.Lucas1918,E.W.Washburn1921}.
As the advanced meniscus rises, the gravitational forces become significant, leading to an equilibrium Jurin's height~\cite{J.Jurin1718}.
Continuous efforts in the study of capillary rise by using experiment, theoretical modeling, and numerical simulation
have been devoted to getting a better understanding of how to manipulate liquid directional transport in tubes or pipes.

It has been found that a variety of factors, such as inner microstructure~\cite{C.Li2019} or geometrical shape of tubes~\cite{V.T.Gurumurthy2022,C.Zhao2021,C.Zhao2022,Y.Chen2009,R.G.Elfego2022} (such as triangular, rectangular and cylindrical tubes), forced wetting~\cite{V.T.Gurumurthy2022}, contact line friction~\cite{T.Andrukh2014,B.K.Primkulov2020,P.Wu2018}, surface roughness of capillary tubes~\cite{Y.Hu2003,S.Girardo2012,P.M.G.Eijo2022}, surfactant concentration~\cite{K.Piroird2011}, and inertial effects~\cite{S.Lunowa2022}, etc, influence the capillary rise.
For example, Li \textit{et al}.~\cite{C.Li2019} proposed a peristome-mimetic inner tube structure with controllable capillary rise and siphon diode behaviors to actuate a larger amount of bulk liquid transfer spontaneously and continuously in an antigravity direction. The capillary invasion of a liquid into an empty tube with various geometries has also been studied extensively~\cite{V.T.Gurumurthy2022,C.Zhao2021,C.Zhao2022,Y.Chen2009,R.G.Elfego2022}.
Through a novel experimental setup, it has been verified that a large fraction of dissipation can take place near the contact line except for the energy dissipated in the fluid bulk during the capillary rise~\cite{B.K.Primkulov2020}.
In addition, theoretical modeling and molecular simulation also provide a useful complement to understanding its mechanism
hidden behind. Mathematical models are constructed to incorporate the dynamic contact angle and inertial effects during the capillary rise of fluids in cylindrical tubes~\cite{S.Lunowa2022}. In Ref.~\cite{N.Fries2008}, based on the momentum balance of a liquid inside a capillary tube, the authors derived an analytic solution for the capillary rise of liquids in a cylindrical tube or a porous medium in terms of height $h$ as a function of time $t$ under several assumptions such as (i) the flow is one-dimensional, (ii) no friction effects is considered, (iii) no inertia or entry effects in the liquid reservoir, and (iv) static contact angle is used.
By carrying out molecular dynamic simulations, Chang \textit{et al}.~\cite{H.Y.Chang2023} investigated the spreading behaviors on graphene sheets and the imbibition process in graphene nanochannels, and found that the total wetting liquid behaves quite differently from the partial one for imbibition dynamics.

Recently, the spontaneous flow of a wetting liquid displacing another immiscible liquid or air in a confined space-usually a capillary tube or porous medium, has attracted great interest~\cite{T.E.Mumley1986I,T.E.Mumley1986II,W.K.Chan2005,M.S.P.Stevar2012,D.Yang2014}.
Continuous interest has been focused on the capillary displacement in a tube prefilled with another immiscible liquid~\cite{B.K.Primkulov2020,P.L.L.Walls2016,J.Andre2020,C.Patrascu2022,P.Wu2018}.
For instance, through a combination of experimentation and modeling, L. L. Walls et al.~\cite{P.L.L.Walls2016} studied the effects of viscosity and gravity on the dynamics of the capillary rise in the viscous regime, they shown that the Lucas-Washburn law is modified when the viscosity of the displaced fluid is comparable to or exceeds the wetting fluid, and the gravity affects the dynamics of the capillary rise not only in the late viscous regime but also in the early viscous regime.
Such capillarity was also found in micro and nanoscale channels~\cite{Y.-T.Cheng2022}. Although these investigations cover various topics, a universal dynamic equation governing the invading process of liquid in an inclined tube is still needed.

In this paper, we construct a model to investigate the capillary invasion of a liquid by another liquid in an inclined capillary tube. Different laws of time-dependent invasion length are obtained under different circumstances, with various factors discussed.

\section{\label{sec:THEORETICAL MODEL}THEORETICAL MODEL}
\begin{figure}[htp!]
  \includegraphics[width=\linewidth,keepaspectratio]{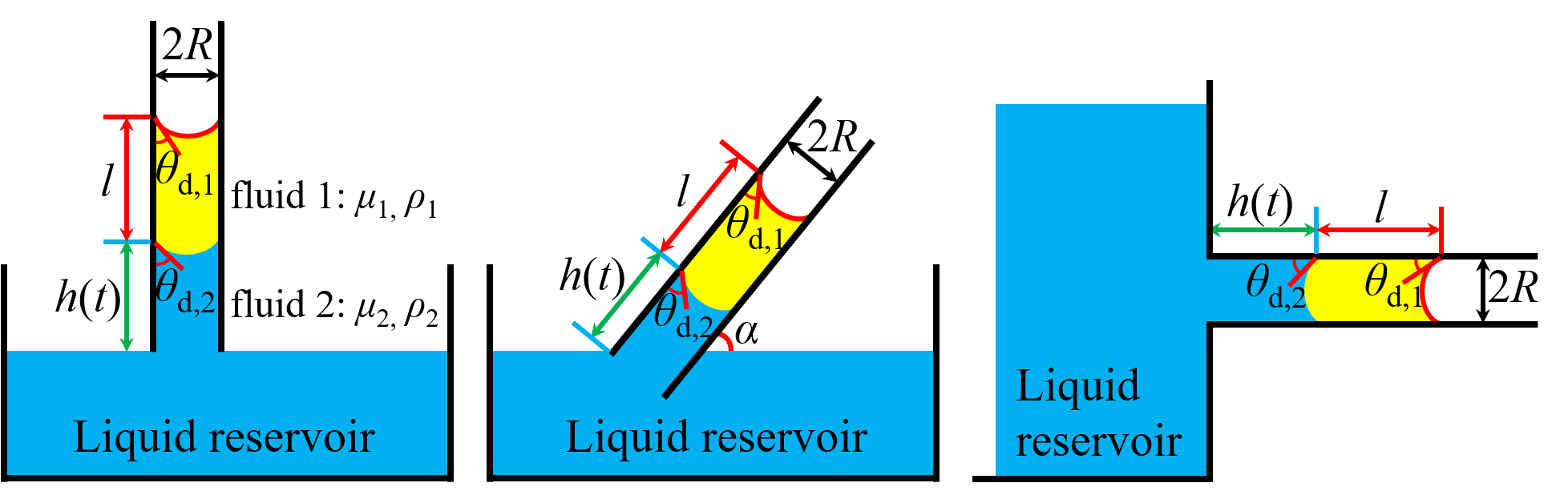}
  \caption{(Color online) The schematic of a liquid imbibition in a capillary tube with a liquid slug of different type. \label{imbibitionprocess}}
\end{figure}

We consider a system sketched from a typical experiment setup, as shown in Fig.~\ref{imbibitionprocess}. Traditionally, the capillary tube was first partially prefilled with the oil with desired length $l$ by exposing one end of the tube to an oil reservoir. Then the end with the slug is put to in contact with a reservoir of water, the water starts flowing into the tube. Thereby, we consider a system of two immiscible liquids in which the viscosity, density, and surface tension of the prefilled oil slug and the later invaded liquid are represented by $\rho_1$, $\mu_1$, $\gamma_1$ and $\rho_2$, $\mu_2$, $\gamma_2$, respectively.
The liquid-liquid displacement system with a liquid slug of different type in a cylindrical tube of radius $R$ [Fig.~\ref{imbibitionprocess}].
Besides viscosity, the pressure loss at the tube entrance also causes energy dissipation (see the Appendix)~\cite{Lorenceau2002}. Here we made a hypotheses that the Landau-Levich-like deposition of thin oil layer is neglected since the thin film dissipation is not considered.
Therefore, the column tends to move due to the capillary force, and be resisted by a combination of gravity, dissipation, and fluid inertia. Using Onsager variational principle, we have the dynamic equation (see the Appendix)
\begin{align}
R^2(\rho_1l+\rho_2h)\ddot{h}+ (1+\frac{3\zeta}{2})R^2\rho_2\dot{h}^2+8(\mu_1l+\mu_2h)\dot{h} +16 R\xi_{\rm eff}\dot{h} \notag\\
+ R^2g(\rho_1l+\rho_2h)\sin\alpha-2 R\gamma_{\rm eff}=0,\label{kineticequationT}
\end{align}
governing the capillary invasion in a tilted tube. Here $\rho_1$ and $l$ are the density and the length of the liquid slug, respectively, $\rho_2$ and $h$ are the density and the imbibition length of the invading fluid, respectively, and $\zeta$ is an coefficient of local resistance to the fluid motion, $\alpha$ is the angle made between the horizontal direction and the tube axis, $\mu_i$ is the dynamic viscosity of the $i$-th liquid, $\xi_{\rm eff}=(\xi_1+\xi_2)/2$ is the effective friction coefficient of the two liquids with $\xi_i=C_i(\mu_i\mu_0)^{1/2}$~\cite{Q.Vo2018PRE,Q.Vo2019}, where $\mu_i$ is the viscosity of $i$-th liquid, $\mu_0$ is the viscosity of the surrounding medium of the liquid, and $C_i$ is a constant depending on roughness and chemical properties of the surface~\cite{A.Carlson2012}. $\gamma_{\rm eff}=\gamma_1\cos\theta_{\rm e,1}+\gamma_{12}\cos\theta_{\rm e,2}$ is the effective surface tension with $\theta_{{\rm e,}i}$ ($i=1, 2$) the equilibrium contact angle of the $i$-th liquid, and $\gamma_1$ and $\gamma_{12}$ the surface tensions of the liquid slug and the slug-invading liquid interface, respectively.
Equation~(\ref{kineticequationT}) indicates an equilibrium imbibition length~\cite{C.Patrascu2022}
\begin{align}
h_{\rm eq}=\frac{2\gamma_{\rm eff}}{\rho_2gR\sin\alpha}-\frac{\rho_1}{\rho_2}l.\label{heq}
\end{align}

In the case of high viscosity, the inertial term can be neglected and the dynamic equation governing the imbibition process in a tilted tube becomes
\begin{align}
\dot{h}=\frac{2 \gamma_{\rm eff}R- R^2g(\rho_1l+\rho_2h)\sin\alpha}{8(\mu_1l+\mu_2h)+16\xi_{\rm eff}R},\label{kineticequationofHighviscosity}
\end{align}
which leads to a solution
\begin{align}
t=-\frac{8\mu_2}{\rho_2gR^2\sin\alpha}h-\tau{\rm ln}\biggl(1+\frac{\rho_2gR\sin\alpha}{\rho_1gRl\sin\alpha-2\gamma_{\rm eff}}h\biggr),\label{SolutionT(h)}
\end{align}
with a typical time scale
\begin{align}
\tau=\frac{8(\mu_1l+2\xi_{\rm eff}R)}{\rho_2gR^2\sin\alpha}+\frac{8\mu_2(2\gamma_{\rm eff}R-\rho_1gR^2l\sin\alpha)}{(\rho_2gR^2\sin\alpha)^2}.\label{tau}
\end{align}
A first-order approximation of Eq.~(\ref{SolutionT(h)}) leads to a linear time-dependent imbibition invasion length
\begin{align}
h=\frac{\gamma_{\rm eff}R}{4(\mu_1l+2\xi_{\rm eff}R)}(1-{\rm Bo}\sin\alpha)t,\label{linear liquid liquid displacement}
\end{align}
for the initial stage in tilted liquid-liquid displacement systems. Here  ${\rm Bo}=(\rho_1gRl)/(2\gamma_{\rm eff})$ is the bond number. If $l=0$ and the contact line dissipation is neglected, Eq.~(\ref{SolutionT(h)}) becomes
\begin{align}
t=-\frac{8\mu_2}{\rho_2gR^2\sin\alpha}\biggl[h+h_{\rm eq}
{\rm ln}\biggl(1-\frac{h}{h_{\rm eq}}\biggr)\biggr],\label{SolutionT(h)l=0}
\end{align}
where $h_{\rm eq}$ is given by Eq.~(\ref{heq}) when $l=0$. If $h\ll\cfrac{2\gamma_{\rm eff}}{\rho_2gR\sin\alpha}$, a second-order approximation of the above equation leads to the well-known Lucas-Washburn law
\begin{align}
h=\biggl(\frac{\gamma_{2}R\cos\theta_{\rm s,2}}{2\mu_2}\biggr)^{1/2}t^{1/2},\label{Lucas-Washburn}
\end{align}
indicating that the invasion length increases in proportion to $t^{1/2}$, where $\gamma_{12}=\gamma_{2}$ and $\theta_{\rm e,2}=\theta_{\rm s,2}$ as $l=0$. The fact that the tilt angle is cancelled out during the approximation also shows that Lucas-Washburn law applies to any tilted tubes.

\section{\label{sec:R&D}RESULTS AND DISCUSSIONS}
\begin{figure}[h!]
  \includegraphics[width=\linewidth,keepaspectratio]{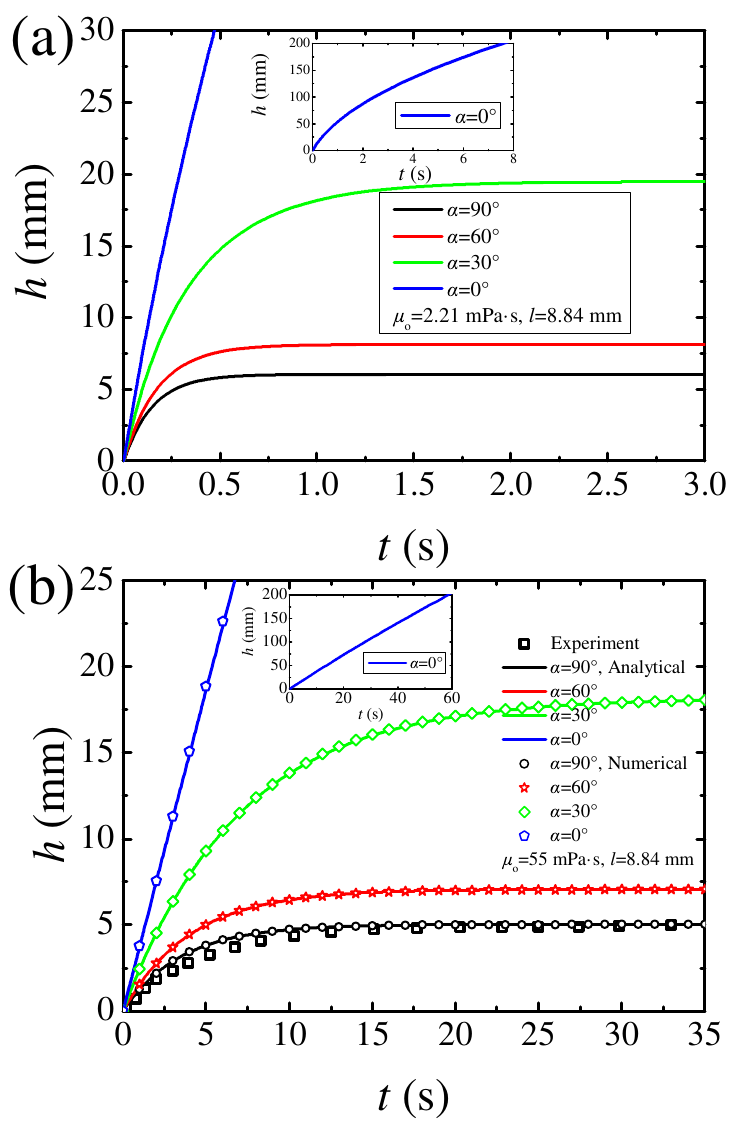}
  \caption{(Color online) The influence of tilt angle on the imbibition of a two fluid flow system with the viscosity of its prefilled fluid set as (a) $\mu_1=\mu_{\rm o}=2.21~{\rm mPa\cdot s}$ and (b) $\mu_1=\mu_{\rm o}=55~{\rm mPa\cdot s}$, respectively, where the tube radius are fixed as $R=0.475~{\rm mm}$. The insets show the invasion length behavior over a long time scale. Under the case $\alpha=90^\circ$, the analytical and numerical results are compared with the experimental data measured by Patrascu \textit{et al.} in Ref.~\cite{C.Patrascu2022}.
  \label{Vary:α}}
\end{figure}
\begin{table*}[htbp]
\caption{Physical properties of imbibing liquids used in this study}
\label{parameters}
\centering
\begin{tabular}{lccccccc}
\toprule
\toprule
Liquids & $\mu~({\rm mPa\cdot s})$ & $\rho~({\rm kg/m^3})$ & $\gamma~({\rm mN/m})$ & $\gamma_{\rm o-w}~({\rm mN/m})$ & $\theta_{\rm o}$ & $\theta_{\rm o-w}$ & References \\ \hline
Silicone oil 1cSt & 0.818 & 818 & 16.9 & $\sim 37.2$ & $-$ & $-$ & \cite{J.Andre2020,Q.Vo2019} \\
Silicone oil 2cSt & 2.21 & 835 & $\sim 20$ & $35$ & $-$ & 60 & \cite{F.Wang2023} \\
Silicone oil 5cSt & 5.2 & 913 & 21 & $\sim 34.2$ & $-$ & $-$ & \cite{R.G.Elfego2022} \\
Silicone oil 50cSt & 48 & 960 & 20.8 & $-$ & $-$ & $-$ & \cite{J.Andre2020} \\
sunflower-seed oil & 55 & 920 & $22\pm2$ & $22\pm2$ & $46.67\pm2.05$ & $37.7\pm3.78$ & \cite{C.Patrascu2022} \\
Silicone oil 100cSt & 97 & 965 & 20.9 & $-$ & $-$ & $-$ & \cite{J.Andre2020,P.L.L.Walls2016} \\
Silicone oil 500cSt & 485 & 970 & 21.1 & $-$ & $-$ & $-$ & \cite{J.Andre2020} \\
Silicone oil 1000cSt & 970 & 970 & $21\sim22$ & $-$ & $-$ & $-$ & \cite{B.K.Primkulov2020,J.Andre2020} \\
Deionized water & 1 & 998 & 72 & $-$ & $-$ & $-$ & \cite{C.Patrascu2022} \\
\bottomrule
\bottomrule
\end{tabular}
\end{table*}
\begin{figure*}[htp]
  \includegraphics[width=\linewidth,keepaspectratio]{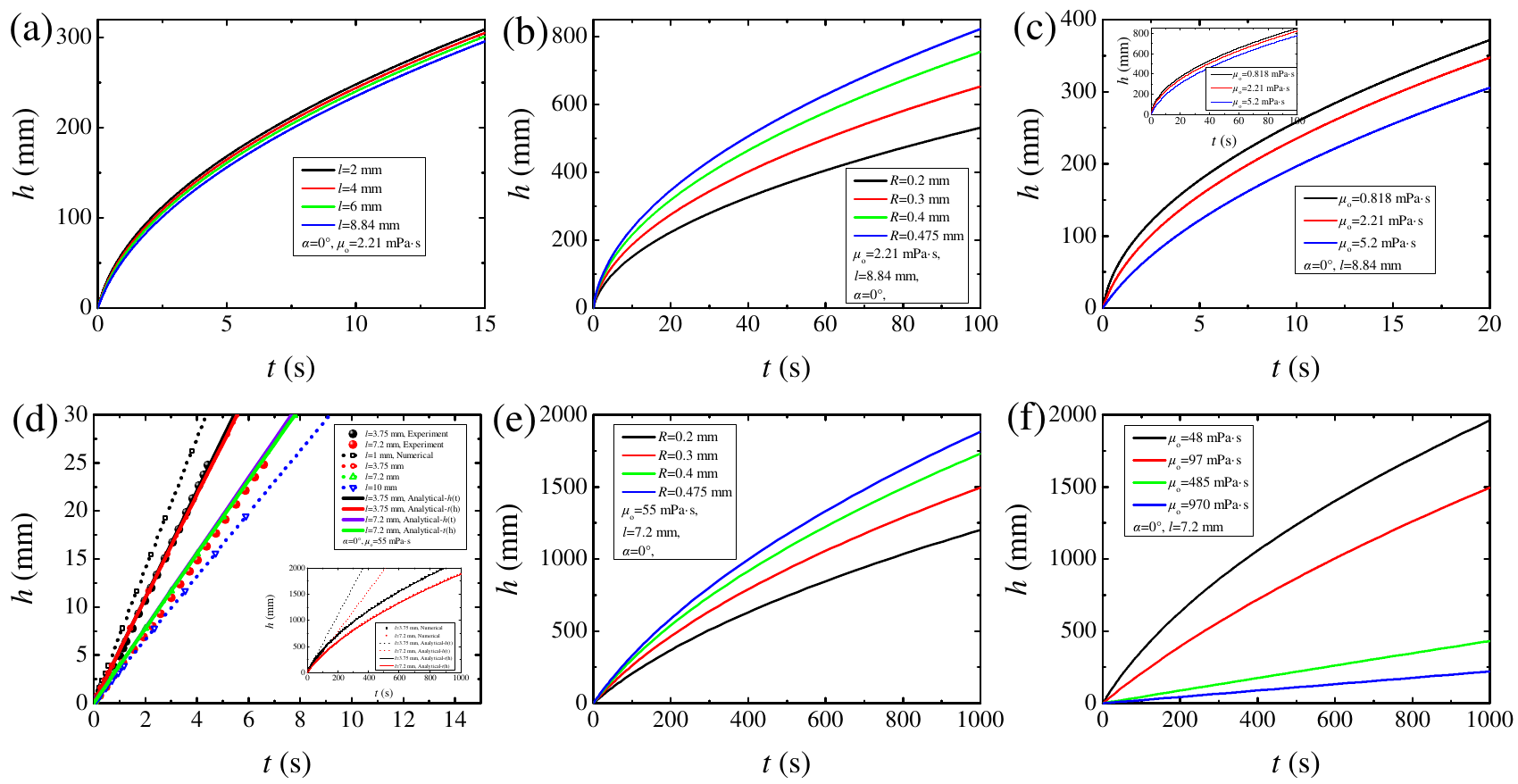}
  \caption{(Color online) The time-dependent invasion length for different (a) prefilled liquid lengths, (b) tube radii, and (c) viscosities of prefilled liquid with $\mu_1=\mu_{\rm o}=2.21~{\rm mPa\cdot s}$. (d), (e) and (f) corresponds to (a), (b) and (c) respectively with $\mu_1=\mu_{\rm o}=55~{\rm mPa\cdot s}$. The cases $l=3.75~{\rm mm}$ and $l=7.2~{\rm mm}$, the analytical and numerical results are compared with the experimental data measured by Patrascu \textit{et al.} in Ref.~\cite{C.Patrascu2022}. \label{Horizontal:High&Low}}
\end{figure*}

The imbibition front $h$ in Eq.~(\ref{kineticequationT}) can be calculated numerically, as shown by the black, the red, the green and the blue curves in Fig.~\ref{Vary:α}(a) with parameters summarized in Table~\ref{parameters}. It is found that the time required for the system to reach the equilibrium increases, and the final equilibrium length decreases with the tilt angle [the green, the red, and the black curves in Fig.~\ref{Vary:α}(a) and (b)], indicating that the gravity plays a significant role in the liquid-liquid displacement.
On the other hand, there is no equilibrium state for a horizontal tube (blue curve). For a small Reynolds number so that $\mu_2h\ll\mu_1l$, the solution to Eq.~(\ref{kineticequationofHighviscosity}) becomes
\begin{align}
h=\frac{2\gamma_{\rm eff}-\rho_1gRl\sin\alpha}{\rho_2gR\sin\alpha}\biggl[1-\exp \biggl(-\frac{\rho_2gR^2\sin\alpha}{8(\mu_1l+2\xi_{\rm eff}R)}t\biggr)\biggr].\label{kineticequationHighviscosity:withdissipationSolution2}
\end{align}
In Fig.~\ref{Vary:α}(b), the black, the red, and the green curves based on the analytical solution Eq.~(\ref{kineticequationHighviscosity:withdissipationSolution2}) agree well with the numerical calculations (the circle, the star, and the diamond dots). The experimental results reported by Patrascu \textit{et al}.~\cite{C.Patrascu2022} (the black square dots on the black curve) also support our model.

Here it is worthwhile to note that for a horizontal tube with $l\neq0$, Eq.~(\ref{SolutionT(h)}) should be replaced by
\begin{align}
&t=\frac{2\mu_2}{\gamma_{\rm eff}R}h^2+\frac{4(\mu_1l+2\xi_{\rm eff}R)}{\gamma_{\rm eff}R}h.\label{SolutionT(h):Horizontal}
\end{align}
Given the imbibition time of interest, the imbibition front can be numerically obtained from this equation, as shown by the blue curves in Fig.~\ref{Vary:α}(b) and its inset. For a small Reynolds number, discarding the nonlinear term of the above equation leads to
\begin{align}
h=\frac{\gamma_{\rm eff}R}{4(\mu_1l+2\xi_{\rm eff}R)}t,\label{SolutionHighviscosity:Horizontal}
\end{align}
corresponding to the linear blue curves in Fig.~\ref{Vary:α}(b). This can also be obtained via a first-order approximation of Eq.~(\ref{kineticequationHighviscosity:withdissipationSolution2}) when $\alpha=0$ or letting $\alpha=0$ in Eq.~(\ref{linear liquid liquid displacement}).

\begin{figure}[htp]
  \includegraphics[width=\linewidth,keepaspectratio]{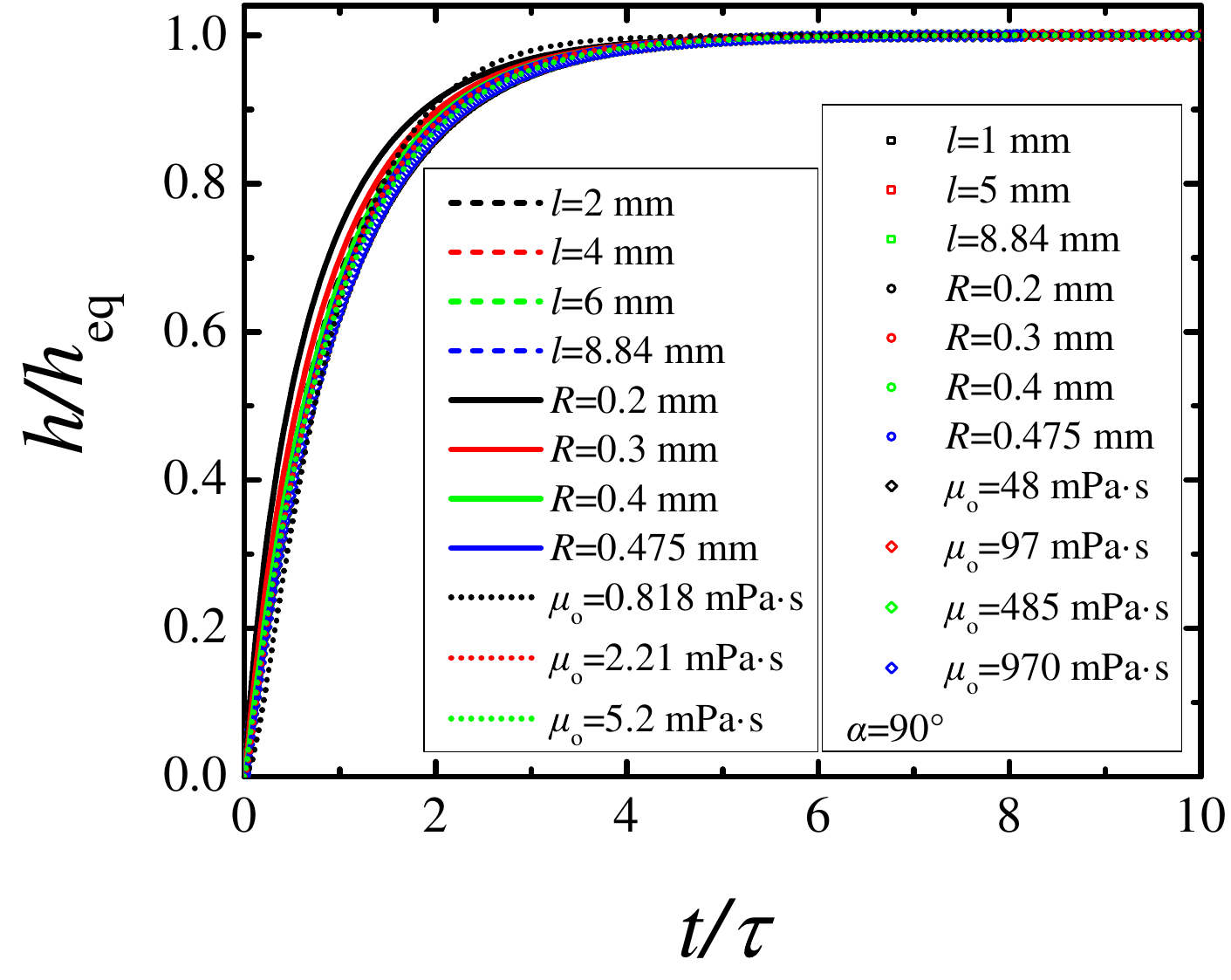}
  \caption{(Color online) Scaled invasion length for different prefilled liquid lengths, tube radii, and viscosities of prefilled liquid, showing a master curve behavior for vertical tube orientation.\label{Vertical:High&Low}}
\end{figure}

In order to investigate how prefilled liquid length $l$, tube radius $R$, and viscosity of prefilled liquid $\mu_{\rm o}$ affect the imbibition dynamics in a horizontal tube, we plot the imbibition length for various controlling factors in Figs.~\ref{Horizontal:High&Low}. If the viscosity of prefilled liquid is small, the invasion length displays a nonlinear behavior [Fig.~\ref{Horizontal:High&Low}(a)]. Otherwise the time-dependent invasion length exhibits a linear feature in short time scale, though a nonlinear variation with time is observed over a long time scale (or $h\ge\mu_1l/\mu_2$) [Fig.~\ref{Horizontal:High&Low}(d) and its inset]. Such nonlinear and linear behaviors correspond to the analytical solutions Eqs.~(\ref{SolutionT(h):Horizontal}) and (\ref{SolutionHighviscosity:Horizontal}), respectively. In Fig.~\ref{Horizontal:High&Low}(d), we can see that our numerical calculations (dot dash curves), analytical solution (black and purple curves), and implicit solution (red and green curves) overlap the experimental observations~\cite{C.Patrascu2022} (the black and the red solid circles), an evidence once again confirming the validity of the present model. It is also shown [Figs.~\ref{Horizontal:High&Low}(a)-(d)] that the invasion speed decreases with the increase of prefilled liquid length $l$ but increases with the increase of tube radius, indicating there exists a trade-off between the dissipation energy and the surface tension energy. The decrease of the Reynolds number enhances the dependence of invasion length on prefilled liquid length [from Fig.~\ref{Horizontal:High&Low}(a) to Fig.~\ref{Horizontal:High&Low}(d)] but slightly change the effect of tube radius [from Fig.~\ref{Horizontal:High&Low}(b) to Fig.~\ref{Horizontal:High&Low}(e)] on the imbibition. Figures~\ref{Horizontal:High&Low}(c) and \ref{Horizontal:High&Low}(f) imply that increasing the viscosity of prefilled fluid leads to the slow down of the invasion dynamics owing to the increase of viscous force.

\begin{figure*}[htp]
  \includegraphics[width=\linewidth,keepaspectratio]{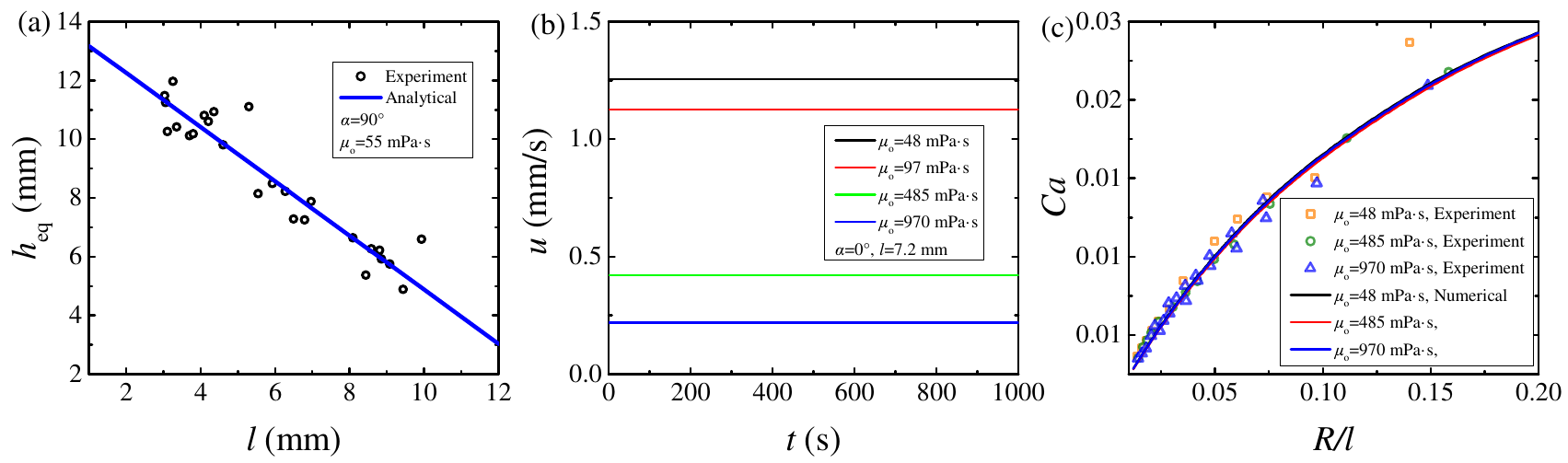}
  \caption{(Color online) (a) Equilibrium length, $h_{\rm eq}$, against the length of prefilled liquid where the analytical results are compared with the experimental results reported by Ref.~\cite{C.Patrascu2022}. (b) Invasion velocity as a function of time. (c) Capillary number $Ca$ against the ratio of the tube radius $R$ to the slug length $l$, where the experimental data come from Ref.~\cite{B.K.Primkulov2020}.\label{Vertical&Horizontal}}
\end{figure*}

Here it should be noted that for a horizontal tube without a prefilled liquid slug, if the dissipation at the contact line is neglected, one obtains the exact solution~\cite{W.K.Chan2005}
\begin{align}
h^2=\frac{\gamma_{2}R\cos\theta_{\rm s,2}}{2\mu_2}t-\frac{\gamma_{2}\rho_2 R^3}{4\mu_2^2}\exp\biggl(-\frac{8\mu_2}{\rho_2R^2}t\biggr),\label{SolutionofHorizontal:Low}
\end{align}
to Eq.~(\ref{kineticequationT}), which also leads to Lucas-Washburn law as $t\to \infty$.

Figure~\ref{Vertical:High&Low} demonstrates the time-dependent invasion length for different prefilled liquid lengths, tube radii, and viscosities of prefilled liquid for the capillary imbibition in vertical tubes, either in small or large Reynolds number cases. It is found that if we use $\tau$ and $h_{\rm eq}$ to scale length time and length, respectively, then all the curves fall into a master one levelling off to 1.

In order to evaluate the accuracy of our model, a comparison is made between the numerical solutions and the experimental observations, as shown in Fig.~\ref{Vertical&Horizontal}. As shown in Fig.~\ref{Vertical&Horizontal}(a), the experimental data can be well captured by Eq.~(\ref{heq}) for a vertical tube~\cite{C.Patrascu2022}. For the liquid-liquid displacement in a horizontal tube with different high viscosities of prefilled liquid, it is also interesting to find that the liquid penetrates the tube at a constant speed [Fig.~\ref{Vertical&Horizontal}(b)]. Here we define a dimensionless capillary number $Ca=\mu_{\rm o}\dot{h}/\gamma_{\rm o}$ to characterize the nominal ratio of viscous force to capillary one, where $\gamma_{\rm o}$ is the surface tension of the oil. As shown in Fig.~\ref{Vertical&Horizontal}(c), it is interesting to find that the experimental results~\cite{B.K.Primkulov2020} and the numerical solutions both collapse onto a master curve, indicating the validity of our model.

\begin{table*}[htbp]
\caption{Formulae of capillary effects}
\label{formulae}
\centering
\begin{tabular}{|c|c|c|c|c|c|}
\toprule
\toprule
\multicolumn{6}{|c|}{$ R^2(\rho_1l+\rho_2h)\ddot{h}+ (1+\frac{3\zeta}{2})R^2\rho_2\dot{h}^2+8(\mu_1l+\mu_2h)\dot{h}+16R\xi_{\rm eff}\dot{h}+ R^2g(\rho_1l+\rho_2h)\sin\alpha
-2 R\gamma_{\rm eff}=0$}\\
\hline
discarding&
\multicolumn{5}{c|}{ $t=-\cfrac{8\mu_2}{\rho_2gR^2\sin\alpha}h-\tau{\rm ln}\biggl(1+\cfrac{\rho_2gR\sin\alpha}{\rho_1gRl\sin\alpha-2\gamma_{\rm eff}}h\biggr)$ (general solution)}\\
\cline{2-6}
inertial&early&$l\neq 0$&$\setminus$ &$\alpha=0$&$\alpha\neq 0$\\
\cline{4-6}
terms&stage& $\xi_{\rm eff}\neq0$&$h\ll\mu_1l/\mu_2$ &$h=\cfrac{\gamma_{\rm eff}R}{4(\mu_1l+2\xi_{\rm eff}R)}t$ & $h=\cfrac{2\gamma_{\rm eff}-\rho_1gRl\sin\alpha}{\rho_2gR\sin\alpha}\biggl[1-\exp \biggl(-\cfrac{\rho_2gR^2\sin\alpha}{8(\mu_1l+2\xi_{\rm eff}R)}t\biggr)\biggr]$\\
\cline{4-6}
& & &$ h\sim\mu_1l/\mu_2$ & $t=\cfrac{2\mu_2}{\gamma_{\rm eff}R}h^2+\cfrac{4(\mu_1l+2\xi_{\rm eff}R)}{\gamma_{\rm eff}R}h$ &general solution \\
\cline{3-6}
&&$\xi_{\rm eff}=0$ & $h\ll\cfrac{2\gamma_{2}\cos\theta_{\rm s,2}}{\rho_2gR\sin\alpha}$ &\multicolumn{2}{c|}{Lucas-Washburn Law~\cite{R.Lucas1918,E.W.Washburn1921}}\\

&& $l=0$&&\multicolumn{2}{c|}{$h=\biggl(\cfrac{\gamma_{2}R\cos\theta_{\rm s,2}}{2\mu_2}\biggr)^{1/2}t^{1/2}$}\\
\cline{2-2}\cline{4-4}\cline{6-6}
&\multicolumn{2}{c|}{} &$h\sim\cfrac{2\gamma_{2}\cos\theta_{\rm s,2}}{\rho_2gR\sin\alpha}$ &&$t=-\cfrac{8\mu_2}{\rho_2gR^2\sin\alpha}\biggl[h+h_{\rm eq}
\ln\biggl(1-\cfrac{h}{h_{\rm eq}}\biggr)\biggr]$\\
\cline{1-1}\cline{4-6}
all terms & \multicolumn{2}{c}{} & \multicolumn{2}{|c|}{$h^2=\cfrac{\gamma_{2}R\cos\theta_{\rm s,2}}{2\mu_2}t-\cfrac{\gamma_2\rho_2 R^3}{4\mu_2^2}\exp\biggl(-\cfrac{8\mu_2}{\rho_2R^2}t\biggr)$~\cite{W.K.Chan2005}}& no analytical solution\\
included&\multicolumn{2}{c}{}&\multicolumn{2}{|c|}{}&\\

\bottomrule
\bottomrule
\end{tabular}
\end{table*}

\section{\label{sec:Conclusion}Conclusion}
In summary, a general theoretical model considering surface tension energy, gravitational energy, kinetic energy, and the dissipation energy (including the bulk dissipation and the dissipation at the contact line) is proposed for a liquid-liquid displacement system by using Onsager variational principle. A couple of new imbibition evolution laws other than Lucas-Washburn type have been found under different circumstances, making our model a comprehensive one, as shown in Table II. The good agreement between our model and experimental results shows that the imbibition dynamics in a capillary tube with a prefilled liquid can be well captured by the evolution equations derived in this paper. The calculation results reveal that increasing the tube's inclination angle, the column's length and the viscosity of prefilled liquid can significantly slow down the imbibition dynamics. The present model can be used to predict the capillary-driven flow in tubular systems, design passive microfluidic devices~\cite{M.Zimmermann2007} with applications in miniature heat pipes to cool electronic components~\cite{L.L.Vasiliev2008}, and pattern biomolecules in microchannels~\cite{E.Delamarche2005} and clinical diagnostics~\cite{C.H.Ahn2004}.

\begin{acknowledgments}
We acknowledge financial support from National Natural Science Foundation of China under Grant No.12147142, No.11974292, and No.12174323.
\end{acknowledgments}

\appendix

\section{\label{sec:Appendixes}Derivation of the dynamic equation Eq.~(\ref{kineticequationT}) in the main text}
For a tilted tube, the total free energy consists of gravitational energy and surface energy, and is given by
\begin{align}
F_{\rm tot}=&(\pi R^2l\rho_1gh+\frac{1}{2}\pi R^2\rho_2gh^2)\sin\alpha \notag\\
&-2\pi Rh(\gamma_1\cos\theta_{\rm d,1}+\gamma_{12}\cos\theta_{\rm d,2}),\label{Ftot_T}
\end{align}
where $\theta_{{\rm d},i}$ ($i=1, 2$) is the dynamic contact angle of the $i$-th liquid. The total energy dissipation of the liquid-liquid displacement system consists of a bulk one
\begin{align}
\Phi_{\rm bulk}&=\frac{\mu_1}{2}\int_0^R2\pi rl\biggl(\frac{\partial u}{\partial r}\biggr)^2 {\rm d}r + \frac{\mu_2}{2}\int_0^R2\pi rh\biggl(\frac{\partial u}{\partial r}\biggr)^2 {\rm d}r \notag\\
&=4\pi(\mu_1l+\mu_2h)\dot{h}^2,\label{bulkdissipation}
\end{align}
under the assumption of a classical Poiseuille flow $u=2(1-r^2/R^2)\dot{h}$, and one due to the friction at the contact line:
\begin{align}
\Phi_{\rm ct}=2\pi R(\xi_1+\xi_2)\dot{h}^2.\label{ctdissipation}
\end{align}
For a liquid flow from a big reservoir into a small capillary tube, it is necessary to take the pressure loss into account, due to an energy loss at the entrance as a result of either a sudden geometry contraction for liquid rise or a sudden geometry expansion for liquid fall. Lorenceau \textit{et al.}~\cite{Lorenceau2002} claimed that such a dissipation of energy depends on the direction of the flow due to the entrance or exit effects. To obtain the rate of energy dissipation due to the entrance effects, we assume that the liquid flows with a velocity $\dot{h}$, then the flux of kinetic energy across the bottom of the capillary tube can be approximated as
\begin{align}
\Delta E_k=\frac{\zeta}{2}(\Delta m)\dot{h}^2,\label{DeltaE}
\end{align}
where $\Delta m=\pi R^2\rho_2\dot{h}(t)\Delta t$ is the amount of mass transferred within time $\Delta t$. Thereby, the inflow rate of kinetic energy is modeled as
\begin{align}
\dot{E}_k=\Phi_{\rm PL}=\frac{\zeta}{2}\pi R^2\rho_2\dot{h}^3,\label{Pressure loss dissipation}
\end{align}
corresponding to the rate of energy dissipation $\Phi_{\rm PL}$ at the tube entrance. Here, $\zeta$ is an coefficient of local resistance to the fluid motion with its sign the same as that of $\dot{h}$. Owing to the rule that the energy dissipation due to entrance/exit effects should be always positive for both liquid rise and liquid fall. As a result, we considered the $\zeta$ with signs in the present model, \textit{i.e.} $\zeta=1$ for the case of liquid rises and $\zeta=-1$ when the liquid falls. Given relationship between the dynamic contact angle and the equilibrium contact angle
\begin{align}
\cos\theta_{{\rm d},i}=\cos\theta_{{\rm e},i}-\frac{\xi_i\dot{h}}{\gamma},\label{theta}
\end{align}
and the velocity
\begin{align}
v=\dot{h}=\frac{\gamma_{\rm lv}(\cos\theta_{\rm e}-\cos\theta_{\rm d})}{\xi},\label{v}
\end{align}
 the change rate of the total free energy Eq.~(\ref{Ftot_T}) can be rewritten as
\begin{align}
\dot{F}_{\rm tot}=&\pi R^2g(\rho_1l+\rho_2h)\dot{h}\sin\alpha \notag\\
&+4\pi R\xi_{\rm eff}(h\ddot{h}+\dot{h}^2)-2\pi R\gamma_{\rm eff}\dot{h}.\label{Ftot_dotT}
\end{align}
This equation, together with Eqs.~(\ref{bulkdissipation}), (\ref{ctdissipation}) and (\ref{Pressure loss dissipation}), leads to a Rayleighian
\begin{align}
\mathcal{R}=&\Phi_{\rm bulk}+\Phi_{\rm ct} +\Phi_{\rm PL}+\dot{F}_{\rm tot}  \notag\\
=&4\pi(\mu_1l+\mu_2h)\dot{h}^2+4\pi R(\xi_1+\xi_2)\dot{h}^2 +\frac{\zeta}{2}\pi R^2\rho_2\dot{h}^3 \notag\\
&+\pi R^2g(\rho_1l+\rho_2h)\dot{h}\sin\alpha+4\pi R\xi_{\rm eff}h\ddot{h}-2\pi R\gamma_{\rm eff}\dot{h}.\label{RayleighianT}
\end{align}
To apply the Onsager variational principle, we still need the kinetic energy of the system, which is given by
\begin{align}
T=\frac{1}{2}\pi R^2(\rho_1l+\rho_2h)\dot{h}^2.
\end{align}
By setting $\partial \mathcal{R}/\partial \dot{h}+\partial (\partial T/\partial \dot{h})/\partial t=0$, the nonlinear dynamic equation Eq.~(1) of the main text is obtained. For a horizontal tube ($\alpha=0$), if the dissipation at the contact line and pressure loss are neglected and there is no prefilled liquid, i.e., $l=0$, Eq.~(\ref{kineticequationT}) of the main text can be reduced to
\begin{align}
\rho_2R^2h\ddot{h}+\rho_2 R^2\dot{h}^2+8\mu_2h\dot{h} -2 R\gamma_{2}\cos\theta_{\rm s,2}=0,\label{kineticequationHTnodissipation}
\end{align}
which has an exact solution Eq.~(\ref{SolutionofHorizontal:Low}) of the main text.

\nocite{*}


\end{document}